\documentclass[12pt,apjfonts,usenatbib]{emulateapj}
\usepackage{natbib}
\input psfig.sty

\begin{document}
\title{Strong dusty bursts of star formation in galaxies falling into the cluster RXJ0152.7-1357.}

\author{
D.~Marcillac\altaffilmark{1},
J.~R.~Rigby\altaffilmark{1},
G.~H.~Rieke\altaffilmark{1} and D.M.~Kelly \altaffilmark{1}}

\altaffiltext{1}{Steward Observatory, University of Arizona, 933
  N. Cherry Avenue, Tucson, AZ~85721}

%



\begin{abstract}

We have observed the cluster RX J0152.7-1357 (z$\sim$0.83) at 24$\mu$m with the
Multiband Imaging Photometer for Spitzer (MIPS). We detected twenty-two sources associated with
spectroscopically confirmed cluster members, while ten more have photometric redshifts 
compatible with membership. Two of the 32 likely cluster members contain obvious active nuclei while the others are associated with dusty star formation.
The median IR-determined star formation rate 
among the remaining galaxies is estimated 
at 22 M$_\odot$$\,$yr$^{-1}$, significantly higher than in previous estimates from optical data. 
Most of the MIR-emitting galaxies also have optical emission lines, but a few do not and hence 
have completely hidden 
bursts of star formation or AGN activity. 

An excess of MIR-emitting galaxies is seen in the cluster in comparison to the field at the same redshift. The MIR cluster members are more associated with previously detected infalling late type galaxies rather than triggered by the ongoing merging of bigger X-ray clumps. 
Rough estimates also show that ram pressure may not be capable of stripping the gas away from cluster outskirt galaxies but it may disturb the gas enough to trigger the star formation activity. Harassment can also play a role if for example these galaxies belong to poor galaxy groups. Thus, bursts of star formation occur in the cluster environment  and could also help consume the galaxy gas content in addition to ram pressure, harassment or galaxy-galaxy strong interactions.

\end{abstract}

\section{Introduction}

Galaxies formed first followed by bigger 
structures such as galaxy clusters and filaments, as demonstrated both by semi-analytic models 
of galaxy formation (combining large cosmological N-body 
simulations with simple, semi-analytic recipes for the baryons, e.g. GALICS, Devriendt \& Guiderdoni, 2000, Hatton 
et al., 2003 ) and also by simulations of structure formation (with gas hydrodynamics, cooling and an empirical 
law for star formation, e.g. RAMSES, Teyssier, 2002). Thus, clusters grow by accreting 
field galaxies, entire galaxy groups, or even entire substructures.
The morphology and spectral properties of the infalling objects are strongly affected by the dense 
cluster environment, and these galaxies rapidly evolve from blue disk dominated galaxies into early type 
galaxies without star formation (E,S0). 

Many mechanisms have been invoked to explain these rapid morphological and spectral changes in cluster galaxies, the 
decrease of the star formation activity of infalling objects, and the lack of star formation in the 
cluster centers. They include interaction with the cluster potential, ram pressure stripping (Gunn \& Gott 1972, Abadi, Moore \& Bower 1999), 
cluster tidal forces (Fujita 1998), 
violent galaxy encounters (Lavery \& Henry, 1994) or rapid galaxy encounters such as galaxy harassment (Moore et al. 1996). However, it is not known whether the infalling objects experience a burst of star formation 
before this process is finally quenched.

Butcher \& Oemler (1978, 1984, BO effect hereafter) found that the fraction of blue galaxies, f$_B$,increased from nearly zero at z $\sim$ 0 to 20\% at z$\sim$0.4. The galaxies responsible for the BO 
effect are associated with star-forming spiral or irregular cluster galaxies 
(Couch et al., 1994, Dressler et al., 1997).   
Spectroscopic studies (Dressler \& Gunn, 1983, Dressler et al., 1985, Dressler \& Gunn, 1992, Dressler 
et al., 1999, Poggianti et al., 1999) confirm the lack of emission-line galaxies among cluster 
members and the decline of strong star formation for moderate-redshift (z=0.35-0.55) 
clusters compared with the moderate-redshift field galaxies. 
Couch et al. (2001) studied H$\alpha$-emitting galaxies in the cluster AC 114 (z=0.32) and concluded 
that star formation is strongly and uniformly suppressed. They found that 
the star formation inferred from H$\alpha$ was always less than 4 M$_{\odot}$/yr, excluding strongly 
star forming systems. In addition, the H$\alpha$ luminosity function was an order of magnitude below 
that observed for field galaxies at the same redshift.
Using the Canadian Network for Observational Cosmology1 (CNOC1) data to provide [OII] emission line
measurements in a sample 
of 15 X-ray luminous clusters at 0.18$\leq$z$\leq$0.55, Balogh et al. (1999) 
confirmed that cluster environments are not responsible for inducing starbursts.
It is still debated whether cluster galaxies experienced a burst of star formation 1-2 Gyrs 
prior to the observations or if the star formation is gradually quenched and whether post starburst cluster members are in excess with respect to 
field galaxies at moderate redshift. Dressler et al. (1999) found that post starburst objects such as 
k+a/a+k galaxies (galaxies without [OII] emission lines but with H$\delta$ 
absorption lines (Dressler et al., 1999), or K+A galaxies (Dressler \& Gunn) 1983), 
are an order of magnitude more frequent in moderate redshift clusters than among field galaxies. 
However, Balogh et al. (1999) did not find an excess of k+a/a+k galaxies in clusters compared with the field.

All the previous studies used optical emission lines to derive star formation properties. However these 
lines can be strongly attenuated by dust and therefore a large part of the star formation 
could be hidden partially or entirely. Using radio data, which is not sensitive to dust extinction, 
Smail et al. (1999) found that a significant number of the apparently post starburst (k+a/a+k) objects in A851 
(z=0.41) are undergoing bursts of star formation that are completely hidden.
Poggianti et al. (1999) have also suggested that dust 
can hide a large part of the star formation in clusters. \cite{PW} showed that 50\% of a sample 
of cluster Very Luminous Infrared Galaxies (log (L$_{IR}$/L$_{\odot}$)$>$11.5 where L$_{IR}$ is the total infrared luminosity in the [8-1000 $\mu$m] range) exhibit e(a) spectra (i.e. 
with O[II] emission lines and EW(H$\delta$)$\ge$4 $\AA$ in the classification from Dressler et al., 1999). This  
emphasizes that these galaxies cannot be interpreted simply as post-starburst galaxies in dust free 
models because they are highly extincted. 

Infrared galaxies are common in the local universe, but extreme objects such as Luminous Infrared Galaxies (LIRGs, with $10^{11}\leq L_{IR}[8-1000 \mu $m$]
($L$_{\sun}) \leq 10^{12}$L$_{\sun}$] ) and Ultra Luminous Infrared Galaxies 
(LIRGs, with L$_{IR}$[8-1000 $\mu $m$]
($L$_{\sun})>10^{12}$L$_{\sun}$] )
are rare (Sanders \& Mirabel, 1996).
They are also found to be extremely rare in local clusters. Bicay \& Giovanelli (1987) studied a sample of 200 FIR sources in seven local clusters and did not find such extreme objects. The survey of several clusters by ISO (see the review from Metcalfe et al., 2005),   Fadda et al. (2000), Biviano et al. (2004) and Metcalfe et al. (2003) confirmed this finding. 

However LIRGs are more common in the distant universe and dominate the infrared luminosity function at
0.5 $<$ z $<$ 1 (e.g., Le Floc'h et al. 2005, P\'erez-Gonz\'alez et al. 2005). 
LIRGs were also found in more distant clusters studied by ISOCAM such as Cl0024+1654 (z=0.39, Coia et al. 2005)  or J1888.16Cl (z=0.56, Duc et al. 2004) showing a significance evolution of the activity of IR sources in clusters.

This paper uses {\it Spitzer}/MIPS data to study infrared-emitting galaxies in the cluster RX J0152.7-1357. 
Our target (section 2) is at z=0.83, where LIRGs play a dominant role in the field. 
The reduction
and analysis of the observations are described in section 3. We present the basic results in section 4,
and discuss them in the context of cluster galaxy evolution in section 5.  
Throughout this paper, we assume H$_o$= 75 km s$^{-1}$
 Mpc$^{-1}$, $\Omega_{\rm matter}$= 0.3 and $\Omega_{\Lambda}= 0.7$.

\section{The Cluster}

The cluster RX J0152.7-1357 was independently discovered in the RDCS survey (Rosati et al., 1998), in 
the SHARC survey ( Romer et al., 2000), and in the WARPS survey (Ebeling et al. 2000).
The rest frame X-ray luminosity within an aperture of 1.5$h^{-1}_{50}$ Mpc is estimated to be L$_X$=(6.8 $\pm$0.6) $\times$ 10$^{44}$ h$_{50}^{-2}$ erg$\,$s$^{-1}$ (Della Ceca et al. 2000). 
Observations from Chandra and XMM show the intracluster medium (ICM) has a complex structure with two 
peaks of X-ray emission, supporting the view of an ongoing merger of these two substructures (Maughan 
et al. 2003, Huo et al. 2004).  

The extensive optical spectroscopic surveys of Demarco et al. (2005) and J{\o}rgensen et al. (2005)
identified respectively 102 and 29 cluster members. They are mainly distributed in the two X-ray substructures and the cluster 
looks like a filament perpendicular to the line of sight according to the distribution of 
these cluster members.
Kodama et al. (2005) used photometric redshifts to find evidence for large scale structures 
around this cluster. 
These structures, one hosting the cluster and oriented NS and another one extending from NE-SSW, were 
spectroscopically confirmed by Tanaka et al. (2006). Blakeslee et al. (2006) showed that there is an offset 
of about 1000 km$\,$s$^{-1}$ between the redshift distributions of late and early type galaxies suggesting 
an infall of late type objects onto the cluster. 

RXJ052 is then a massive, dynamically young, and unvirialized cluster. 
Because merging of substructures and infalling are detected, it is 
an interesting target to study whether dust enshrouded star formation can be induced in a 
cluster environment or if this environment only quenches star formation.

\section{Observations, data reduction, source detection and catalogs}

\subsection{MIR data}
Deep 24 $\mu$m observations were obtained in August 2004 with MIPS.  
We observed a 10' $\times$ 5' field for a total of 3.6 ksec per sky pixel. The data were reduced
with the MIPS Data Analysis Tool (Gordon et al., 2005). 
The final mosaic has an interpolated pixel size of 1$\farcs$245.
An expected 1 $\sigma$ rms photon noise level of 12 $\mu$Jy should be reached in this exposure. 
However we performed aperture photometries at several positions where no objects were detected and 
estimated a detection level at 1 $\sigma$ rms of 19 $\mu$Jy in the inner 5'$\times$5' field. The latter takes into account both the photon noise and the confusion limit. 
We concentrated on this region hereafter because of the lack of spectroscopic redshifts outside it. 
The astrometry of the field was checked with the ESO star catalog and NED and is better than 2". 
Point source extraction and photometry were performed using DAOPHOT (Stetson, 1987) as described 
in Papovich et al. (2004), since the sources we are studying are unresolved.
We detected 201 sources with $fl_{24}$ $\ge$ 60 $\mu$Jy among which 153 are $\ge$ 83 $\mu$Jy.
 
\subsection{Optical data}
In addition, the cluster was observed with the HST ACS in November 2002 in the F625W, F775W, F850LP 
bandpasses (PI : H. Ford). 
We
downloaded the flat--fielded images from the HST archive, cross-correlated
to measure the alignment offsets not accounted for by the world coordinate
system, and then mosaicked the images using MultiDrizzle version 2.5.

\subsection{X-ray data}
We downloaded the available Chandra data (PI : H. Ebeling) for RXJ0152.7-1357 from the
archive, which included 34~ks of good data (excluding flares).  We reduced
the data using Ciao 3.2, following the CXC threads.  
We also created a point-source catalog from the Chandra data.  To do so,
we reduced each ACS-I chip separately, creating exposure--corrected images
in the standard bands (0.5--8 keV, 0.5--2 keV, 2--8 keV, 4--8 keV).  To
identify sources, we ran the Ciao task Wavdetect on each chip, in each
band, using the spatial wavelet scales used by Brandt et al. (2001), which are optimized for point source detection.  
For the wavdetect significance threshold of $10^{-7}$, there are likely to
be 0.4 spurious sources in our catalog.
For each source, we performed photometry in each band, on the
exposure--corrected image, using a circular aperture whose radius is the
$90\%$ encircled energy radius at that detector position and energy band.  
The sky annulus had an inner radius equal to the $90\%$ encircled energy
radius, and an outer radius twice as large.  We converted the
exposure--corrected photons to ergs~cm$^{-2}~s^{-1}$ by multiplying by the
flux-weighted energy of the band, and applying a $10\%$ aperture
correction.

\subsection{MIR catalog of cluster members}

We cross-correlated the spectroscopically confirmed galaxy members and non-members from Demarco et al. 
(2005), J{\o}rgensen et al.(2005), and Tanaka et al. (2006) with the MIPS objects, using a 2'' search radius. 
This positional tolerance accounts for the MIPS positional accuracy and the fact that the brightest optical 
emission can be shifted from the MIR peaks (e.g., as in the Antennae galaxy, Mirabel et al., 1998). Twenty-two 
members (18 from the Demarco list, 3 more from the Tanaka list and 1 more from the J{\o}rgensen list) were associated 
with unique MIR sources, while one association was rejected because the optical counterpart is not unique.
An additional 17 MIR sources appear to be background or foreground objects.
Using only spectroscopic redshifts allowed us to associate 19\% of the MIPS sample with a unique redshift, 
which emphasizes the need to also use photometric redshifts. 

Kodama et al (2005) performed deep observations (reaching M$_V$* + 4) on RXJ0152.7-1357 in the VRi'z' Suprime-Cam bands and derived photometric redshifts using the photometric redshift code presented in Kodama, Bell \& Bower (1999).
Tanaka et al. (2006) compared $z_{spec}$ and $z_{zphot}$ showing that  
$\Delta$z=$z_{zphot}$ - $z_{spec}$ = 0.01 $\pm$ 0.02 \footnote{median value with 1-$\sigma$ error bar.} for the 200 galaxies with a spectroscopic redshift in the 0.76-0.88 range\footnote{$\Delta$z=0.01 $\pm$ 0.02 for the subsample of galaxies with a photometric redshift in this range and a spectroscopic redshift.}, showing the good accuracy of their photometric redshifts. However photometric redshift estimation sometimes suffers from small numbers of strong outliers ($\Delta$z$\ge$0.4). These wrong redshift estimations do not have a strong influence when studying statistical properties of a large number of galaxies and are taken into account in error bars, but can be important when using a small subsample of galaxies from a bigger sample of photometric redshifts. We have checked in the final sample of spectroscopic redshifts the probability that a  galaxy with a spectroscopic redshift out of the 0.76-0.88 range, has a $z_{zphot}$ falling in this range. It appears that of 151 such sources, only 7\% have a photometric redshift in the 0.77-0.88 range.  
In addition to the outliers, the spectroscopic data used to compare with the photometric redshifts may be biased against red cluster members, for which photometric estimates are known to be more efficient. Bluer galaxies (including many MIPS detections) may have less-confident redshifts; Tanaka et al (2006) showed that their photometric redshifts do not work well for very blue galaxies at z$\sim$0.8. 

We have therefore checked the accuracy of the photometric redshifts on the MIPS sources. Taking into account the subsample of these sources  with a spectroscopic redshift, we estimated the accuracy of the photometric redshifts to be $\Delta$z$_{MIPS}$=+0.01$^{+0.02}_{-0.03}$, which shows that the photometric redshifts using the VRi'z' Suprime-Cam bands work well for MIPS sources.
However they do not work for the two most luminous objects in the MIR, which are AGN (detailed in subsection 4.1) since $\Delta$z=z$_{spec}$-z$_{phot}$=-0.08 and -0.58 for them \footnote{As it is shown in section 4.1, the other MIPS sources do not show any obvious evidence of AGN activiy. }

We cross-correlated the 162 other MIR sources with the photometric redshift list from Tanaka et al. (2006). We associate 99 out of the 162 sources with a unique optical counterpart, 
among which 10 have a photometric redshift in the 0.80-0.87 range and may be cluster members.
We were not able to associate a unique optical counterpart for 45 of the MIR detections, but only four of these cases among all the possible optical counterparts have photometric redshifts 
in the [0.76-0.88] range. As a consequence, we have omitted these MIR objects from our study because they have a very low probability of belonging to the cluster. No optical counterparts on either the ACS image or the Tanaka 
et al. (2006) images were found for the other sources.

Two subsamples were defined for the purpose of this paper : the $z_{spec}$ subsample, composed of 
the 22 spectroscopically confirmed 24$\mu$m-detected cluster members, and the $z_{zphot}$ sample
with 
the ten likely 24$\mu$m members that have only a photometric redshift in the 0.80-0.87 range \footnote{Tanaka et al. (2006) chose the 0.76-0.88 range to study the large scale structure close to this cluster. Given our interest
in the cluster alone, we will use a range of 0.80-0.87 in this paper.}.
Studying cluster membership with photometric redshifts is dangerous given typical photometric redshift errors : assuming a cluster velocity dispersion V$_{disp}$ to be $\sim$1320 km$\,$s$^{-1}$ (Girardi et al., 2005), {\it{dz}}=V$_{disp}$/c=0.005, which is a bit smaller than $\Delta$z$_{MIPS}$. 
As a consequence most of this study is based on the spectroscopic subsample while we only use the second list to estimate upper limits in Sections 4 and 5.


Most of the first subsample (19/22) have a 24$\mu$m flux density $\ge$ 83 $\mu$Jy, 
which is the 80\% completeness limit estimated by Papovich 
et al. (2004) for the MIPS Deep survey. This bias is mainly due to the difficulties in associating sources 
with 60$\mu$Jy$\le$ $fl_{24}$ $\le$ 83$\mu$Jy with a unique counterpart.
More than half the members of the second subsample (7/10) are brighter than 83 $\mu$Jy.

Some MIPS sources may have an optical counterpart fainter than r$_{AB}$$\sim$26.5 (r$_{Vega} \sim$ 26.3) which is the magnitude limit of the catalog used for deriving the photometric redshift. As a consequence we may have missed them in our two catalogs.
However Le Floc'h et al (2005) showed that most of the MIR sources with fl$_{24}$ $\ge$ 83 $\mu$Jy and z$\sim$ 0.8 are expected to be associated with sources with R$_{Vega}<$24 while fainter sources are usually at higher redshift. Even if we study sources down to 60 $\mu$Jy, we do not expect to miss a significant number of MIPS sources since only 18 of 202 are not associated with optical counterparts. In fact, some of these sources are outside the ACS or Subaru fields, which can easily explain the lack of optical counterparts.

We also cross-correlated the X-ray catalog with the two MIR 
cluster member catalogs with a radius of 3" to take into account of astrometric uncertainties.

\section{Results}

\subsection{IR Galaxies dominated by an active nucleus}
The two most luminous galaxies at 24 $\mu$m are also detected in the X-ray, and 
their optical spectra show broad MgII($\lambda$2798) (Demarco et al., 2005), unambiguous 
signatures of AGN activity. One is associated with a very disturbed isolated galaxy while the other 
belongs to a compact group of at least 5 disturbed and strongly interacting galaxies within a 5''$\times$ 3'' 
region.

None of the other MIR detected galaxies are identified as a single X-ray source, although we 
cannot exclude the presence of obscured, dusty AGN. 
In the field, galaxies detected in the MIR range are predominantly associated with dusty star forming 
galaxies rather than AGN (Fadda et al., 2002). 
At lower redshift, La Franca et al (2004) showed that at least 25 \% of the MIR sources detected in the ELAIS-S1 field are associated with an AGN according to their optical spectra and stressed this percentage is a lower limit since some objects whose spectra do not show signs of AGN activity are detected in the X-ray. However, as emphasized in Oliver \& Pozzi (2005) the presence of an AGN does not imply that it dominates the total infrared luminosity and using SED modelling, Rowan-Robinson (2004) showed that the total infrared luminosity is dominated by AGN activity for only 11 \% of the total ELAIS sample and 29 \% in SWIRE (Rowan-Robinson, 2005). 

As a consequence, even if such studies have not been conducted for galaxy clusters, 
we will assume that the other galaxies of our sample are dominated by dusty starbursts.

\subsection{Star formation rate and infrared luminosity}

The mid-infrared emission of a LIRG is dominated by a combination of a continuum 
associated with very small grains \citep[VSG]{DBP90} stochastically heated and fluctuating 
around temperatures of a few hundred degrees, and broad emission features at 3.3, 6.2, 7.7, 
8.6, 11.3 and 12.7\,$\mu$m associated with aromatic molecules 
loosely called {\it{Polycyclic Aromatic Hydrocarbon}} \citep[PAH]{PL}.
Silicate grains are responsible for absorption features at 9.7 and 18\,$\mu$m. 

\begin{figure}
       \resizebox{\hsize}{!}{\includegraphics{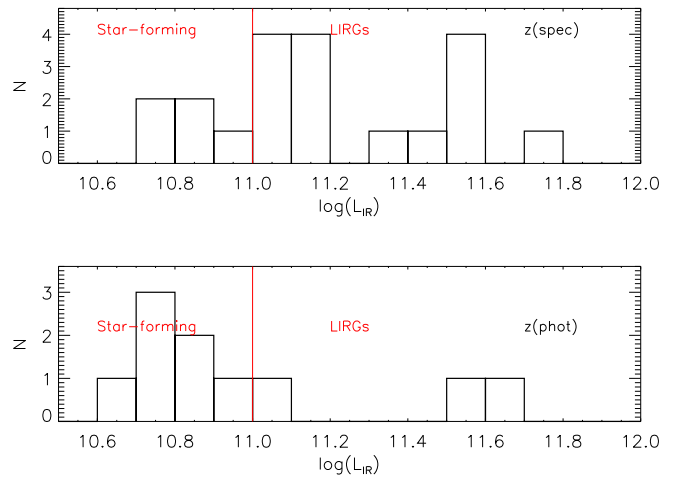}}
      \caption {Distribution of L$_{IR}$ for MIR cluster members with a spectroscopic redshift (top) and a photometric redshift (bottom). }
\label{LIR}
\end{figure}

Thus, at z$\sim$ 0.8, the 24 $\mu$m emission is associated with a combination of the 12 $\mu$m PAH feature and the 
VSG continuum in the galaxy rest frame. The luminosities at 12 and 15$\mu$m correlate with the L$_{IR}$ in the local 
universe \citep{CE} and at least up to z$\sim$1.2 (Marcillac et al., 2006b). 
We therefore estimate the infrared luminosity of each galaxy using the spectral energy distribution (SED) 
from the Dale \& Helou (2002) library as explained in Marcillac et al. (2006b). Each of the 
template SEDs is redshifted to the distance of the observed source. Then, the SED whose MIR flux density 
is the closest to the observed one is used to derive L$_{IR}$ after it is normalized 
to the exact observed flux density. Marcillac et al. (2006b) compared L$_{IR}$ estimated using 24$\mu$m 
data to L$_{IR}$ estimated using radio data for LIRGs and found a scatter of 40\%  between 
the two estimates. 
Fig~\ref{LIR} shows a histogram of L$_{IR}$ for our sample of galaxies : 15 objects have a total infrared 
luminosity above $10^{11} L_{\odot}$ and are LIRGs, while 5 lie just below this range (the 
AGN have been excluded from Fig~\ref{LIR}). The histogram of sources selected with photometric redshifts 
is mainly composed of lower luminosity star-forming galaxies (L$_{IR}<$10$^{11}$L$_{\sun}$).

The star formation rates in units of solar masses per year, SFR(IR) were derived using the relation of Kennicutt (1998) :
\begin{equation}
SFR(IR,M_{\odot}/yr)=1.71 \times10^{-10}(L_{IR}/L_{\odot}). 
\end{equation}
The resulting SFRs(IR) range between 9 and 86 M$_\odot$/yr with a median SFR = 22$_{-10}^{+42}$ M$_\odot$/yr, significantly 
higher than found with optical lines such as [OII] (SFR[OII] hereafter). 
For example, we find that the median SFR(IR) of the emission-line galaxies studied in J{\o}rgensen et al. (2005) is 22   
M$_\odot$/yr, which is about 10 times higher than their optical SFR([OII]). 
These high values of SFR(IR) suggest a large amount of dust-embedded star formation. 

There are a number of additional reasons for this difference beyond the strong dust attenuation
at the blue wavelength of the [OII] line. Besides extinction, 
[OII] is also very sensitive to 
metallicity and the ionization state of the gas (Kennicutt 1998); in addition, this line can be polluted 
by AGN activity. H$\alpha$ is known to be 
a better star formation indicator, particularly when corrected for attenuation and underlying 
stellar absorption (Liang et al., 2004, Flores et al., 2004 for distant LIRGs, 
Hopkins et al., 2003 for local ones), but it is redshifted into the near infrared window, polluted by strong sky 
lines and often impossible to detect. Thus, [OII] often is the only 
feasible line to measure at the redshift of the cluster. 

\subsection{Spatial distribution of the MIR-detected galaxies}

We next discuss the spatial distribution of the MIR cluster members in comparison to the X-ray emission, its 
substructures and the distribution of the other cluster members.
Figure~\ref{PL0} shows the ACS image of the 3'$\times$3' cluster center. The blue contours 
show the Chandra X-ray emission and trace the ICM and point sources while the red ones show the MIPS sources. 

\subsubsection{Star formation activity in the inner X-ray clumps.}

Among 9 MIPS sources located in the X-ray contours, only one is possibly associated with a cluster member 
with $z_{phot}$$\sim$0.8. 
The other photometric redshifts do not fall in the 0.76-0.88 range.
This is in agreement with Blakeslee et al. (2006) who found that there are no blue cluster members (brighter 
than i$_{775}$=23) in the area of high galaxy number density. Similarly Demarco et al. (2005), J{\o}rgensen et 
al (2005) and Homeier et al. (2006) found that emission line galaxies are absent from this area. 
This lack of star formation activity may be attributed to interaction 
with the ICM, tidal stripping or galaxy harassment that have previously removed the gas from the galaxies.
As a consequence, these objects must have formed their stars at earlier epochs.

\subsubsection{Star formation activity between the two merging clumps.}

Figure~\ref{PL0} shows the region between the two X-ray clumps, thought to be in 
the process of merging. As detailed in Section 2, interactions with the hot ICM between the two
structures could either trigger or quench star formation. 
Annis (1994) found that a sample of clusters with z$\sim$0.33 exhibits a blue fraction, $f_B$, 
ranging from 0 to 50\%. 
He suggested this tendency toward a high (but also highly variable) blue fraction 
is consistent with episodic star formation in clusters. One of his hypotheses 
is that many galaxies in the cluster undergo a starburst simultaneously due to the merging of a subcluster 
clump into a main cluster.

Eleven sources with f(24) $\ge$ 60 $\mu$Jy are detected in the region between the X-ray clumps.
However, one of these objects has a spectroscopic redshift incompatible with being a cluster member,
while the remaining ten have z$_{phot}$ out of the range 0.77-0.88.
Therefore, dusty star formation is not detected between the X-ray emission peaks. The merging
region is not the site of intense star formation, at least at the level detectable at 24$\mu$m.

\subsubsection{Star formation activity in the cluster outskirts}

The spatial distribution of the outer cluster members is given in Figure~\ref{PL}, where the galaxy cluster 
members are indicated by blue triangles, LIRGs by red ones and lower-luminosity 
star-forming objects by yellow ones. The 
spectroscopically confirmed members are indicated by filled symbols and the photometric ones by 
empty ones. The black circles identify late-type galaxies (see section 4.4.2 for more details). 
The two AGNs are indicated as stars. 
Since the cluster looks more like a filament than a virialized symmetric structure, we studied the distribution of galaxies and MIR detected ones relative to the filament 
defined in Figure~\ref{PL} instead of the radial distance as it is currently studied.

Figure~\ref{DP} shows the distribution of the distance to the filament for the non-MIR-detected members 
(plain line), the MIR-detected $z_{spec}$ sample (dashed lines)  and the MIR-detected sample with 
membership from either a $z_{spec}$ or $z_{phot}$.   
There is a lack of MIR-detected galaxies in the cluster center; MIR-detected galaxies tend to lie in 
the outskirts of the filament 
and seem to be less concentrated than the non-24$\mu$m-detected cluster members. 

A Kolmogorov-Smirnov test on these two subsamples of galaxies reveals that the probability that these two 
distributions come from the same parent is about 16 \% for the $z_{spec}$ sample. and about 14 \% for the $z_{spec}$ and  $z_{phot}$ sample.
MIR sources appear sometimes to have a slightly different spatial distribution with respect to the other galaxy cluster members : Coia et al. (2005) found that 15 $\mu$m detections, mainly associated with LIRGs, are less 
concentrated than the non-MIR-detected members in Cl0024+1654, while Biviano et al. (2004) found no significance difference between MIR and galaxy cluster members in A2218 (z=0.175).

\begin{figure*}

      \caption {The 24 micron (red) and Chandra (blue) contours overlaid on an ACS image of the cluster center. The Chandra contour levels are associated with 9.0$\times$10$^{-18}$, 
2.1$\times$10$^{-17}$, 
3.3$\times$10$^{-17}$, 
4.6$\times$10$^{-17}$, 
5.8$\times$10$^{-17}$ 
erg$\, $s$^{-1} \, $cm$^{-2}$ arcsec$^{-2}$ while the MIPS ones are associated with 
4.0$\times$10$^{-7}$, 
1.1$\times$10$^{-6}$, 
2.8$\times$10$^{-6}$, 
7.5$\times$10$^{-6}$, 
2.0$\times$10$^{-5}$ 
Jy$\,$arcsec$^{-2}$.}
\label{PL0}
\end{figure*}

\begin{figure*}
       \resizebox{\hsize}{!}{\includegraphics{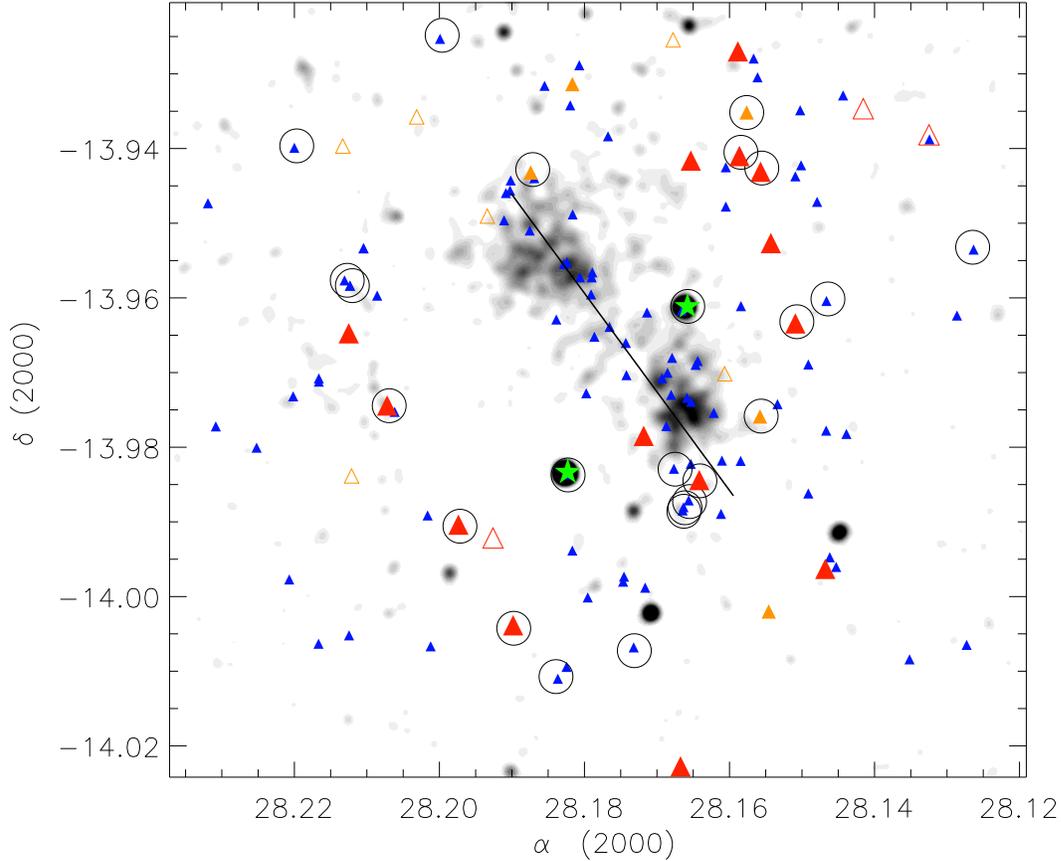}}
      \caption {Distribution of galaxy cluster members (plain symbols) or candidate cluster galaxies for IR sources (e.g. with a photometric redshift compatible with being a cluster member, empty symbols) with respect to the X-ray emission (black and white colors). 
 The different symbols discriminate galaxy cluster members without star formation (blue triangles), LIRGs (red triangles), 
and star forming galaxies (orange triangles). 
The two AGNs are indicated as stars. The black circles are 
associated with late type galaxies (see section 4.4.2 for more details). The line defines the X-ray filament as we use it to discuss the position of galaxies in the cluster   }
\label{PL}
\end{figure*}

\begin{figure}
       \resizebox{\hsize}{!}{\includegraphics{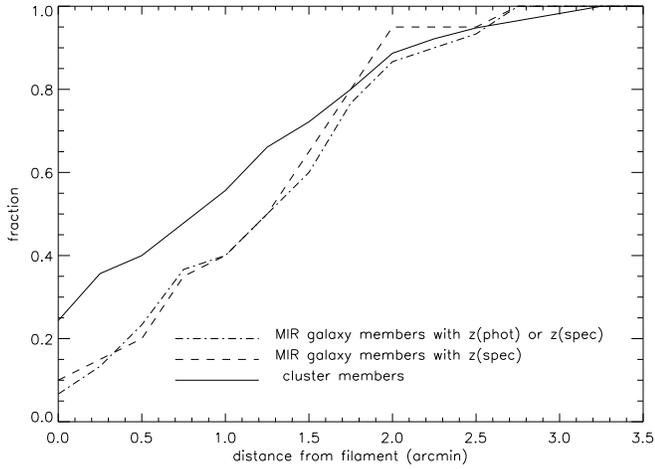}}
      \caption {Distribution of the filament distance of all cluster members (plain line), the spectroscopically confirmed MIR member (dashed lines) and the MIR detected cluster members or with 0.80$\leq$$z_{phot}$$\leq$0.87 (dash-dotted line).}
\label{DP}
\end{figure}

\subsection{MIR-detected galaxy properties}

We now examine the optical properties of the MIR galaxies in comparison with those of the other member galaxies.  

\subsubsection{Spectral Properties}

\begin{table*}
\caption{Morphology of galaxy cluster members with $fl_{24}$ $\ge$60 $\mu$Jy} 
\begin{center}
\label{S}
\begin{tabular}{ccccccc}
\hline
\hline
Class        &\multicolumn{3}{c}{Type \& Definition}  &\multicolumn{2}{c}{MIR member sample}   &all galaxy cluster \\    

             &   EW[OII]3727    &   EW H$\delta$  &    comments                                              & fl$_{24} \ge$ 60. $\mu$Jy & fl$_{24} \ge$ 83. $\mu$Jy &  \\ 
\hline
\hline
e(n)$^{a}$        &   ...            &    ...          &    AGN                      &2  &  2 &3\\
e(a)$^{a}$        &   Yes            &    $\ge$4       &                             &2  &  2 &2\\
e(b)$^{a}$        &   $\ge$40        &    ...          &                             &2  &  1 &14\\
e(c)$^{a}$        &   Yes,$\le$40    &    $<$4       &                               &6  &  6 &13\\
e$^{a}$           &   Yes            &    ?            &    low S/N      &1  &  1 &1\\
k+[OII]+a$^{b}$   &                  &                 &                             &1  &  1 &2\\
k$^{a}$           &   Absent         &    $<$3       &                               &3  &  1 &58\\
k+a$^{a}$         &   Absent         &    3-8          &                             &1  &  1 &8\\
a+k $^{a}$        &   Absent         &    $\ge$8       &                             &0  &  0&1\\



\hline
\hline
\end{tabular}
\end{center}

\noindent{\footnotesize {\it Comments} : (a) Classification from Dressler et al. (1999). (b) Classification from Demarco et al. (2005) : these galaxies exhibit a k+a spectrum with [OII] superposed. Following Dressler et al. (1999), they should be classified as e(c) galaxies. However they are redder than the other e(c) galaxies and as red as k type ones.} 
\end{table*}

Cluster galaxies can be divided into several categories that link their spectral properties to 
their recent star formation history. Dressler et al. (1999) summarize these differences based 
on the [OII]3727 emission line and H$\delta$ absorption one (see also Table~\ref{S}): 

\begin{itemize}
\item passive galaxies (k type) : their spectra do not exhibit emission lines and EW(H$\delta$) $<$ 3 $\AA$.
\item  post-starburst (k+a/a+k type) :  no emission lines are detected in their spectra. ``a+k'' galaxy spectra exhibit stronger Balmer absorption lines (EW(H$\delta$) $>$ 8 $\AA$) than k+a ones ( 3 $\AA$$<$EW(H$\delta$)$<$ 8 $\AA$).
\item Emission-line galaxies (e(a)/e(b)/e(c)/e(n)) : e(a) galaxies exhibit emission lines and strong Balmer absorption lines (EW(H$\delta$$>$ 4 $\AA$); e(b) galaxies are associated with spectra with EW([OII])$>$ 40 $\AA$ while e(c) galaxies are associated with objects with moderate emission lines (EW([OII]$<$40$\AA$) and moderate absorption lines (EW(H$\delta$)$<$4 $\AA$). e(n) are sometimes associated with AGN.
\end{itemize}
Demarco et al. (2005) classified the members of RXJ0152.7-1357, 
following Dressler et al. (1999); they associated 18/22 of the spectroscopic subsample of 
MIR sources with a spectral type. 
Table~\ref{S} presents their spectral types for MIR and non-IR galaxies and confirms that MIR emission 
is mainly associated with emission-line galaxies. While 100\% (e.g. 2/2) and $\sim$50\% (e.g. 6/13) of e(a) and e(c) 
galaxies are detected respectively in the MIR, less than 15\% (e.g. 2/14) of the e(b) ones are seen there. Only two e(b) 
galaxies of this subsample are detected: one with a flux at 24 $\mu$m of 85 $\mu$Jy and the other one 
even fainter. 
These differences can be explained by the effects of extinction in the MIR-bright galaxies. 
Since galaxies are classified as type e(b) if they have a strong [OII]3727 emission line 
(equivalent width greater than 40 $\AA$), this type must be less affected by extinction and thus less 
frequently detected in the MIR.

Another perspective is that $\sim$77\% \footnote{The object defined by a 'k+[OII]+a' type as been associated with e(c) objects since it should be classified as e(c) using the Dressler et al. (1999) classification. } of the sample is made up of galaxies whose burst of star formation/AGN activity 
is only partly hidden. Only four MIR objects exhibit spectra of passive galaxies (k) or post-starbust 
(k+a) galaxies without emission lines, which means that the burst of star formation is completely hidden (see Smail et al. 1999). 
The object with a ``k'' spectral classification can be explained by an old passive stellar population 
where a burst of star formation has just started, since the absorption lines are not affected by 
a very young burst of star formation (Marcillac et al. 2006a). 
It is not clear how to interpret the star forming history of the k+a galaxy; it could 
be a galaxy that has experienced a burst of star formation in the previous 2 Gyr and that is now
experiencing a new one, or where the previous one is continuing. However, it could also be explained by very dusty AGN activity.   

\subsubsection{Morphology}

Galaxy morphology is an important clue to galaxy evolution, and is complementary to 
spectral classification.
Postman et al. (2005) visually classified all the galaxies in the ACS RX J0152.7-1357 field with i$_{775}$ $<$23.5 into three broad morphological categories: E (elliptical; -5 $\leq$T$\leq$-3), SO 
(lenticular; -2 $\leq$T$\leq$0), and Sp+Irr (spiral+irregular;1$\leq$T$\leq$10). We separate the last 
category into Spiral (Sa-Sd, 1$\leq$T$\leq$7) and Irregular (Sdm, Sm, Im,  8 $\leq$T$\leq$10) 
categories. We were able to associate a morphology with 16/22 of the spectroscopic subsample of MIR galaxies. 
 Table~\ref{morpho} presents the morphological type for the MIR and non-MIR galaxies, and emphasizes the large proportion of spiral galaxies among MIR sources. The MIR  sources are associated with mid-to-later type star-forming disk galaxies and not the evolved early-type cluster galaxies, showing that their morphology has not yet been strongly affected by the cluster environment. 
\begin{table*}
\centering
\begin{tabular}{cccc}
\hline
\hline
  Type      & fl$_{24} \ge$ 60. $\mu$Jy & fl$_{24} \ge$ 83. $\mu$Jy& galaxy cluster$^{a}$  \\    
\hline
E              &0  & 0  &41\\
S0             &3  & 2  &35\\
Sp (Sa-Sd)     &9  & 8  &22\\
Irr (Sdm-Irr)  &2  & 2  &7\\
AGN            &2  & 2  &2\\
\hline
\hline
\end{tabular}
\caption{Morphology of galaxy cluster members with 24 $\mu$m with $fl_{24}$ $\ge$60 $\mu$Jy} 
\label{morpho}
\noindent{\footnotesize {\it Comments} : (a) Data from Blakeslee et al. (2006)} 
\end{table*} 

\subsubsection{Color-color diagram}
\begin{figure}
       \resizebox{\hsize}{!}{\includegraphics{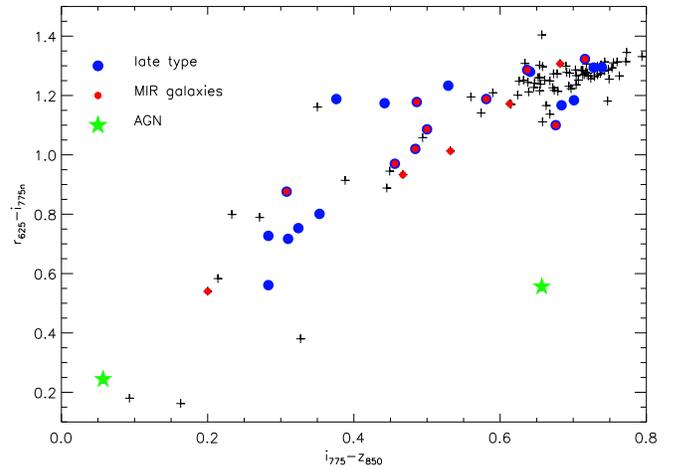}}
      \caption {Color-color diagram for galaxy cluster members.  
The different symbols indicate late type galaxies (blue plain/empty circles), MIR galaxies (red plain circles) or galaxy cluster members not associated to late type or MIR emitting objects (crosses). The morphology classification is detailed in Postman et al. (2005), while the color data are from Blakeslee et al. (2006).}
\label{CC}
\end{figure}

We now examine the optical colors (i$_{775}$-
z$_{850}$ and 
r$_{625}$-
i$_{775}$) of the MIR-detected galaxies compared with those of the other cluster members. We 
cross-correlated our sample of cluster members and non members with the optical measurements published by 
Blakeslee et al. (2006) to derive the color of both IR and non-IR galaxies: colors were 
available for 16 IR galaxies.  The
MIR galaxies studied here are actively star forming systems and one can expect that their optical 
colors may be affected by dusty star formation: they could be redder if the extinction is stronger. 
They also might be bluer if stars with longer evolutionary timescales than O or B stars had time to escape their parental giant 
molecular clouds. After escape, they are expected to be less extincted according to differential 
dust attenuation models such as Charlot \& Fall (2000). 
Figure~\ref{CC} compares the cluster members (dark crosses), late-type galaxies (blue circles), MIPS sources (red circles) in a color-color diagram. The colors of the IR galaxies do not 
differ significantly from those of the other late-type galaxies in this cluster.

\subsubsection{Redshift distribution}

\begin{figure}
       \resizebox{\hsize}{!}{\includegraphics{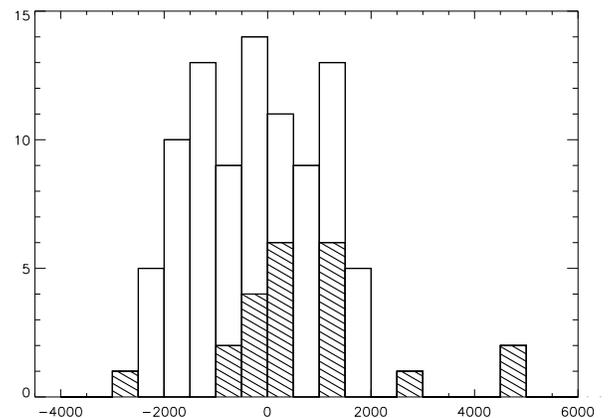}}
      \caption {Histogram of the velocity dispersion relative to the cluster barycenter for non-MIR galaxy cluster members and MIR detected cluster members (hatched). }
\label{V}
\end{figure}

Figure~\ref{V} shows a histogram of the velocity dispersion of all the non MIR galaxy members in 
addition to that of the MIR detected galaxies with a spectroscopic redshift. The velocity 
dispersion of the whole cluster has been estimated to be $\sim$1320 km/s (Demarco et al., 2005, Girardi et al., 2005).
The median redshift of the IR galaxies is estimated to be 0.840$_{-0.003}^{+0.006}$\footnote{value with 
1 $\sigma$ error bar.}, while the whole sample has a median redshift of 0.837.

A Kolmogorov-Smirnov test on the MIR sample and the total set of galaxy members shows that the probability that these two 
distributions come from the same parent is about 6.0 \%. 

This difference in the two distributions is virtually identical to the difference found by Blakeslee et al. (2006) between the 
early-type (median redshift $\sim$ 0.833) and late-type cluster members (median redshift $\sim$0.840 ), suggesting that the former are falling onto the latter. 
Thus, the MIR emitting galaxies probably belong to the infalling group of late-type galaxies. 


\section{Discussion}

\subsection{Comparison between MIR cluster members and MIR field galaxies.}

We have shown that the 24$\mu$m-detected cluster members are located in the outskirts of the cluster, 
avoiding both the ICM region and where the subclump merging occurs.
A large fraction of these sources have with spectra with emission lines.
About 60\% of the MIR-detected galaxies for which we have morphologies are associated with late-type objects 
and 8/11 \footnote{7 are classified late type galaxies while 1 is classified as Sdm.}of the 
LIRGs with a morphological classification in this sample are late-type objects. 
These ratios are higher than the fraction of LIRGs associated 
with spiral objects found by Melbourne 
et al. (2005) at 0.61$<$z$<$1.00 (although our results are subject to small-number
statistical uncertainties). However, our ratios agree well with the results of Bell et al. (2005), 
who studied the properties of 24$\mu$m-detected galaxies with 0.65 $<$z$<$ 0.75. 

Homeier et al.(2006) found that late-type cluster galaxies have similar properties, such as 
physical sizes and disk lengths, as the late-type field galaxies, though they exhibit 
redder optical colors. 
Because of the redder colors, Homeier et al. (2006) claimed that these galaxies 
cannot be a pristine infalling population. However the MIR-detected galaxies have the same colors as the other late-type cluster galaxies. As a consequence, the burst of star formation does not affect the optical colors and some other mechanism must play a role to explain the late type redder colors.

\subsection{ Comparison of the dusty star formation in local clusters and field galaxies : is the RXJ0152 environment triggering bursts of star formation ?}

Are infrared luminous galaxies in excess in this cluster with respect to field galaxies or other galaxy clusters ?
In the CDFS, Le Floc'h et al. (2005) found 2966 objects at 24 $\mu$m with a flux density greater than 83 $\mu$Jy for 
775 arcmin$^2$, e.g. 3.8 sources per arcmin$^2$. We found 6.1 sources per arcmin$^2$ above this limit 
in the cluster field. This difference can be explained by cosmic variance and by the fact that the CDFS is known to be 40\% and 90\% underdense at  respectively z$\sim$ 0.75$\pm$0.1 and 0.85$\pm$0.1 (Wolf et al., 2003).

In the CDFS, 203 objects are detected with 0.81$<$z$<$0.87 and flux density at 24$\mu$m $\ge$ 83$\mu$Jy. 
Taking into account the 80\% completeness detection limit at 83 $\mu$Jy for Spitzer deep surveys 
(Papovich et al. 2004), assuming a redshift completeness limit of 100\% for sources with z$\sim$ 0.8 
(see Figure 3 of Le Floc'h et al., 2005), and a maximum underdensity of 90 \%, an upper limit of 0.6 sources per arcmin$^2$ is expected.

The cluster studied here contains 19 sources detected at 24$\mu$m with a spectroscopic redshift in the same redshift and luminosity range, 
which represents 0.8 sources/arcmin$^2$. 
However, this density is a lower limit since we did not use the galaxies with 
photometric redshifts: 7 such galaxies are detected above 83 $\mu$Jy, which sets an upper limit 
of 1 source/arcmin$^2$.
It should be noted that no completeness correction has been applied to the 24$\mu$m detections in the cluster where it has been applied in the field,
since it is 
difficult to estimate how many outlying MIR galaxies are members of the 
cluster.
Such sources would not change significantly the results, but would only 
increase the excess of 24 $\mu$m sources. Thus, an excess of 24$\mu$m sources is 
detected in the cluster with respect to field galaxies located at the same redshift.
RXJ0152.7-1357 is the first cluster that has been shown to have an excess of LIRGs and star-forming MIR 
galaxies (with fl$_{24}$ $>$ 83 $\mu$Jy) compared with the field.

A small excess of MIR star forming galaxies has been detected for the rich cluster at z=0.181 by Fadda et al. (2000), but these 
IR galaxies are lower-luminosity star-forming objects. 
Using ISOCAM, Coia et al. (2005) found 6 LIRGs and 10 objects with with L$_{IR}$/L$_{\sun} \geq 9 \, $10$^{10}$
in Cl 0024+1654 (z$\sim$0.39) from a survey of 38 
arcmin$^2$, plus two other galaxies without spectroscopic redshifts that may be LIRG cluster members. 
As a consequence, the LIRG density\footnote {taking into account all the sources with L$_{IR}$/L$_{\sun} \geq 9 \, $10$^{10}$.} in this cluster is $\sim$ 0.30 sources arcmin$^{-2}$.We used the 80 \% completeness limit which is estimated in Metcalfe et al. (2003) for A370 since A370 and Cl 0024+1654 have an equivalent exposure at 15 $\mu$m and are located at a comparable redshift, and applied a 20 \% completeness correction  to the previous value. 
Elbaz et al. (1999) present number counts for ISOCAM surveys at 15$\mu$m. Using their integral 
counts above 320 $\mu$Jy (which is the LIRG limit at z = 0.39 according to the Dale \& Helou SEDs), one can expect a 
density of sources at least as luminous as LIRGs ranging from 0.39 to 0.66 sources arcmin$^{-2}$. The range 
arises from cosmic variance between the ultra deep ROSAT and FIRBACK fields. 
Then no excess of MIR sources is detected in Cl 0024+1654. This is in agreement with Geach et al. (2006) who studied  the outskirt of Cl 0024+1654 with MIPS and did not find a clear excess of MIPS sources when comparing number counts in the cluster field to the MIPS deep field number counts except in its center part where a slight excess is suggested. After having removed the bulk of the background field contamination, they find an excess of MIR sources in Cl 0024 but they did not find an excess in MS 0451 (z$\sim$0.55).     
As a consequence, the deficit/excess of star formation with respect to field galaxies shows that we cannot simply expect more MIR in the cluster because the galaxy density and gravitational potential is much higher. The presence
of LIRGs  and star formation may be more connected to the dynamical state of the cluster and mechanisms such as merging of small substructures and infall of galaxies or poor groups that increase the star formation activity.    
 

\subsection{Dusty star formation activity and growth of structures.}

As we saw in section 4, the merging of the two X-ray substructures does not lead to an excess of star formation at the interface. 
This can be explained if most of the star formation occurred at an earlier stage of the merging process. 
This explanation would require that the merging timescale for the 
two substructures be longer 
than the triggering timescale for star formation and its duration. As a consequence, these galaxies may 
have consumed a large amount of their available gas, and the ram pressure they are currently experiencing is stripping 
away the rest. However six galaxies studied by Demarco et al. (2005) that lie between the X-ray 
clumps exhibit a ``k'' spectral type. As a consequence, they probably did not experience a burst of star formation in 
the last two Gyr since it would have affected the Balmer absorption lines (Marcillac et al., 2006b).

The LIRGs and star forming galaxies tend to be located in the outskirts of the cluster.
The arguments by Blakeslee et al. (2006) that the late-type galaxies are infalling on the cluster are detailed in 
section 4.5.1, where we also showed that the MIR-detected galaxies 
are associated with the infalling group.
We do not know the MIR-detected galaxy motion perpendicular to the line of sight,  but if the galaxies had already 
passed through the dense ICM and cluster core, they would have lost all their gas content, preventing  
them from experiencing IR bright phases. 
This suggests the only possible motion perpendicular to the line of sight is to fall on the cluster for the 
first time. 

The infall of a gas rich isolated galaxy or small group of galaxies that contains some gas rich galaxies is thus more likely to be responsible for the LIRG phase rather than the infalling or merging of bigger substructures. These larger structures tend to be X-ray detected and are where galaxy evolution was influenced long ago by the cluster environment. Star formation has already been quenched and is no longer affected by the process of merging of big substructures. The intense episode of
star formation in infalling galaxies helps consume their gas. In the field the duration of such an episode was estimated 
to be 0.1 Gyr and such a phase could consume about 5 $\times$ 10$^9$ M$_{\odot}$ of gas (Franceschini et al., 2003,  
Marcillac et al., 2006b). However, the burst duration could be shorter in the cluster environnment since it is more hostile than the field. 

\subsection{Probable origins of the burst of star formation.}

What causes the bursts of star formation in these galaxies?
It is not known what triggers IR-bright phases such as LIRGs in field galaxies. 
In the local universe, mergers or interactions between galaxies are usually suggested, but single galaxies 
can also experience a LIRG phase (Ishida \& Sanders, 2001). In clusters, other mechanisms such as interactions 
with the ICM, the cluster potential violent encounters between galaxies, or harassment 
have also been suggested for triggering LIRGs (see introduction for references).

In this cluster, the LIRG mechanism cannot be associated with violent encounters between galaxies, since it would lead to an excess of peculiar 
or  irregular galaxies rather than the spirals that dominate the population.

Ram pressure is a unique aspect of the cluster environment. Following Gunn \& Gott (1972), 
the ram pressure experienced by a galaxy is 
$P_{rp}$ = $\rho$V$^2$
where $\rho$ is the ICM density and V the galaxy relative velocity.
It can be expressed as follows :  
\begin{equation}
P_{rp} = 2 \times10^{-12}\times 
(\frac{v}{1000(km \, s^{-1})})^2 \times
\frac{\rho}{10^{-4} (cm^3 s^{-1})}.
\end{equation}
Assuming a relative velocity of 1000 km.s$^{-1}$ and an ICM density of about $10^{-4}$ ($cm^3$ $s^{-1}$), 
which is what is experienced by galaxies in the cluster outskirts,
\begin{equation}
P_{rp} = 2 \times10^{-12}  dyn \, cm^{-2}.
\end{equation}
The gravitational pressure  defined by $P_g$=2$\pi$G $\sigma_{star} \sigma_{gas}$ can be expressed as follows, 
where {\it{x}} is the total galaxy to gas mass ratio.
\begin{equation}
P_g=2\times \pi G x  
\frac{M_{star}^2}{r_{gal}^4}.   
\end{equation}
or
\begin{equation}
P_g \sim 2 \times 10^{-7} x \times \frac{M^2 (10^{10}M_{\odot})}{r^4(kpc)}.  
\end{equation}

We will assume the cluster members have the same mass and gas properties as field LIRGs in the same redshift 
and $L_{IR}$ range, since we saw that the LIRGs are associated with late-type galaxies and the latter have similar size 
properties as the field galaxies (see section 5.1).
The incompleteness of the photometric data for our sample of galaxies
prevents determining their stellar masses.  Franceschini et al. (2003)
computed stellar masses from UV-optical-NIR spectra of IR luminous galaxies detected at 15 $\mu$m.
From their Table 6, a
total of 6 LIRGs have a spectroscopic redshift between $z$=0.6 and
1.0, and 2 more have a photometric redshift in this range. After
converting Franceschini's values to H$_o$= 75 km s$^{-1}$ Mpc$^{-1}$,
we find a median stellar mass of 6$\times$10$^{10}$ M$_{\odot}$ for
the galaxies with spectroscopic redshifts. Including the less
robust photometric redshifts does not change this value. 

The HI and $H_2$ gas masses of local LIRGs have been estimated in Sanders \& Mirabel (1996) and 
Mirabel \& Sanders (1985,1988). Sanders \& Mirabel (1985) found a tight correlation between L$_{FIR}$(40-400 $\mu$m) 
and M$_{H_2}$ for local bright radio spirals with $L_{FIR}$ ranging between $10^{10}$-$10^{12}$ $L_{\odot}$. 
Assuming the Soifer et al. (1991) relation L$_{IR}$=1.91L$_{FIR}$, the median M$_{H_2}$ mass for galaxies in 
the same L$_{IR}$ range is 4$\times$$10^9$ M$_{\odot}$. 
Mirabel \& Sanders (1988) studied the HI gas mass of the most luminous far infrared galaxies in the 
local universe. They found no correlation between $L_{FIR}$ and M$_{HI}$. Using their sample, the 
median M$_{HI}$ for galaxies in the same L$_{IR}$ range is 6$\times$$10^9$ M$_{\odot}$ even if a large 
dispersion is observed. A value of 0.16 can therefore be estimated for {\it x}, and

\begin{equation}
P_g \sim   \,\, 10^{-11} dyn \, cm^{-2}.
\end{equation}

\noindent
Since this value exceeds our rough estimate of P$_{rp}$, the ram pressure may not be adequate to 
strip the gas away from the galaxy plane. However, its effects could be strong enough to induce star forming 
activity by disturbing the gas; it could also make the gas lose its angular momentum and induce AGN activity by making it move to the central massive black holes or produce a nuclear starburst. 

As the infall progresses, the ICM density $\rho$ will become stronger while {\it{x}} will decrease due both to ram pressure and star 
formation activity. As a result, the gas will be stripped away and the star formation activity finally quenched.

However, our previous estimates are very crude and harassment can also account for the burst of star 
formation especially if the galaxy belongs to a small group of objects being accreted by the cluster.
The major point is that it is quite plausible that galaxies experience a strong burst of star formation when falling 
into a cluster before different mechanisms transform them into early-type galaxies. Then, the star formation and the ram pressure stripping in addition to harassment will eventually remove the gas from the galaxies.

\section {Summary and Conclusion}
We used MIPS at 24$\mu$m to study the dusty star formation in the cluster RX J0152.7-1357, located at z$\sim$0.83.
This cluster is an interesting target since its complicated X-ray image suggests two 
substructures are merging while it is also suspected to be undergoing an infall of late-type galaxies. 

Our main results are summarized as follows :

$\bullet$ There are 22 MIR-detected spectroscopically confirmed cluster members, among which are 
two AGNs, 15 LIRGs and 5 lower-luminosity star-forming galaxies. In addition 
10 other MIR sources have a photometric redshift compatible with membership. The median SFR(IR) is 
estimated to be 22 M$_{\odot}$/yr, much larger than the optically estimated star formation rate
and showing that a large amount of the star formation is hidden by dust.

$\bullet$ However, the majority of the MIR-detected galaxies are associated with emission-line late type 
galaxies, so dust only partially hides the star formation. Only four objects exhibit a high rate 
of star formation without 
optical emission lines. There is no 
significant difference between late-type galaxies with and without 24$\mu$m emission in the optical 
color/color diagram. 
  
$\bullet$ An excess of MIR-emitting galaxies is seen in this cluster with respect to field galaxies at the same 
redshift. This is the first time such an excess of LIRGs/star forming objects in 
a cluster has been shown unambiguously. This behavior can be associated directly with the state of the 
cluster rather than just with the high galaxy density of the cluster.

$\bullet$ The MIR galaxies seem to be mostly associated with the infall of late-type galaxies rather than 
the merging of bigger X-ray substructures, showing that a burst of star formation can occur during galaxy infall. 
These sources could either be isolated or belong to small groups being accreted by the cluster.
Thus, some cluster environments may be responsible for a burst of star formation. This behavior seems also to be related to the infalling system properties such as its size and dynamical state.
The ram pressure may not be adequate to strip away the gas but interaction between the galaxy and the ICM may induce perturbations 
leading to a burst of star formation. This perturbation could also be associated with harassment when galaxies 
encounter the greater galaxy concentration in the cluster.
In any case, this burst of star formation can help quench future star formation by 
consuming the available gas and enhancing the gas loss by ram pressure stripping.

\acknowledgments
This work is based on observations made with the {\em Spitzer} Space
Telescope, which is operated by the Jet Propulsion Laboratory,
California Institute of Technology under a contract with NASA
(contract number \#1407).  Support for this work was provided by NASA
through an award issued by JPL/Caltech (contract number \#1255094).  
The authors would like to thank the anonymous referee for helpful comments. D.M. is grateful to Emeric Le Floc'h and Casey Papovich for assistance in data analysis, photometry process and helpful discussions, Masayuki Tanaka for providing photometric redshifts and helpful comments, to Joannah L. Hinz and Christopher N. A. Willmer for stimulating discussions related to this paper and David Elbaz for providing ISOCAM number counts.





\nocite{*}
\bibliography{bib}

\newpage
\newpage

\end{document}